\documentclass[a4paper,iopams]{jpconf}

\usepackage{graphicx}

\def\openone{\leavevmode\hbox{\small1\kern-3.3pt\normalsize1}}

\begin{document}
\title{Fermion-fermion and boson-boson amplitudes: surprising similarities}

\author{Valeri V. Dvoeglazov}

\address{Universidad de Zacatecas, 
Apartado Postal 636, Suc. UAZ,
Zacatecas 98062, Zac., M\'exico}

\ead{valeri@planck.reduaz.mx}

\begin{abstract}
Amplitudes for fermion-fermion, boson-boson and fermion-boson interactions are
calculated in the second order of perturbation theory in
the Lo\-ba\-chev\-sky space.
An essential ingredient of the model is the Weinberg's
$2(2j+1)-$ component formalism for describing a particle
of spin $j$. The boson-boson amplitude is then compared
with the two-fermion amplitude obtained long ago by Skachkov
on the basis of the Hamiltonian formulation of quantum field
theory on the mass hyperboloid, $p_0^2 -{\bf p}^2=M^2$, proposed
by Kadyshevsky. The pa\-ra\-met\-ri\-za\-tion of the amplitudes by means
of the momentum transfer in the Lo\-ba\-chev\-sky space leads to
same spin structures in the expressions of $T-$ matrices
for the fermion case and the boson case. However,  certain
differences are found. Possible physical applications are discussed.
\end{abstract}

The scattering amplitude for the two-fermion interaction had been
obtained in the 3-momentum Lobachevsky space~\cite{Kadysh} in the second order of perturbation
theory  long ago~[2a,Eq.(31)]:
\begin{eqnarray}\label{eq:TF}
\lefteqn{T^{(2)}_V ({\bf k} (-) {\bf p}, {\bf p}) =
-g_v^2 \frac{4m^2}{\mu^2 +4 \mbox{{\bf \ae}}^{\,2}} -
4g_v^2\frac{(\bf{\sigma}_1 \mbox{{\bf \ae}})(\bf{\sigma}_2
\mbox{{\bf \ae}}) - (\bf{\sigma}_1 \bf{\sigma}_2)
\mbox{{\bf \ae}}^2}{\mu^2 +4\mbox {{\bf \ae}}^{\,2}} -\nonumber}\\
&-& {8g_v^2 p_0 \mbox {\ae}_0 \over m^2}\,
\frac{i\bf{\sigma}_1 [{\bf p} \times \mbox{{\bf \ae}} ] +i\bf{\sigma}_2
[{\bf p} \times \mbox{{\bf \ae}} ]}{\mu^2 +4 \mbox{{\bf \ae}}^{\,2}} -
{8g_v^2 \over m^2}\,\frac{p_0^2 \mbox{\ae}_0^2 +2p_0 \mbox{\ae}_0 ({\bf p}
\cdot \mbox{{\bf \ae}}) - m^4}{\mu^2 +4\mbox{{\bf \ae}}^{\,2}} -\nonumber\\
&-& \frac{8g_v^2}{m^2}\,\frac{(\bf{\sigma}_1 {\bf p})
(\bf{\sigma}_1 \mbox{{\bf \ae}}) (\bf{\sigma}_2 {\bf p})
(\bf{\sigma}_2 \mbox{{\bf \ae}})}{\mu^2 +4\mbox{{\bf \ae}}^{\,2}}\quad,
\end{eqnarray}
$g_v$ is the coupling constant. The additional term (the last one) has usually {\it not} been taken into account 
in the earlier Breit-like calculations of two-fermion interactions.
This consideration is based on use of the formalism
of separation of the Wigner rotations and parametrization
of currents by means of the Pauli-Lubanski vector, developed long ago~\cite{Chesh}. The quantities
$$\mbox{\ae}_0 = \sqrt{\frac{m(\Delta_0 +m)}{2}}\quad,\quad
\mbox{{\bf \ae}} = {\bf n}_\Delta \sqrt{\frac{m(\Delta_0 -m)}{2}}$$
are the components of  the 4-vector of a momentum half-transfer.
This concept is closely connected with a notion of the half-velocity
of a particle~\cite{Chernik}.
The 4-vector $\Delta_{\mu}$:
\begin{eqnarray}
\bf{\Delta} &=& \bf{\Lambda}^{-1}_{{\bf p}} {\bf k}
= {\bf k} (-) {\bf p} = {\bf k}
-\frac{{\bf p}}{m} (k_0 - \frac{{\bf k}\cdot
{\bf p}}{p_0 +m})\,,\\
\Delta_0 &=& (\Lambda^{-1}_{p} k)_0 = (k_0 p_0
-{\bf k}\cdot{\bf p})/m = \sqrt{m^2\,+ \,\bf{\Delta}^2}
\end{eqnarray}
can be regarded as the momentum transfer vector in
the Lobachevsky space instead of the vector ${\bf q}= {\bf k} - {\bf p}$ in the Euclidean space.\footnote{I keep  a notation and  a terminology
of ref.~\cite{Skachkov}.  In such an approach all particles (even in the intermediate states) are on the mass shell (but, spurious particles present). The technique
of construction of the Wigner matrices $D^J (A)$  can be found in
ref.~\cite[p.51,70,English edition]{Novozh}.
In general,
for each particle in interaction
one should understand under 4-momenta $p^\mu_i$ and $k^\mu_i$\, $(i=1,2)$
their covariant generalizations, $\breve{p}^\mu_i$, $\breve{k}^\mu_i$,
{\it e.g.}, refs.~\cite{Chesh,Faustov,Dvoegl2}:
$$\breve{{\bf k}} = (\bf{\Lambda}_{{\cal P}}^{-1} {\bf k}) =
{\bf k} - \frac{\bf{{\cal P}}}{\sqrt{{\cal P}^2}} \left ( k_0 -
\frac{\bf{{\cal P}}\cdot{\bf k}}{ {\cal P}_0 + \sqrt{{\cal P}^2}}
\right )\quad,$$
$$\breve{k}_0 = (\Lambda_{{\cal P}}^{-1} k)_0 =
\sqrt{m^2 +\breve{{\bf k}}^{\,\,2}},$$
with ${\cal P}= p_1 +p_2$,
$\Lambda_{{\cal P}}^{-1} {\cal P} = ({\cal M},\, \bf{0})$.
However, we omit the circles above the momenta
in the following, because  in the case under consideration
we do not miss physical information
if we use the corresponding quantities in c.m.s.,
${\bf p}_1 = -{\bf p}_2 = {\bf p}$ and ${\bf k}_1 =-{\bf k}_2 = {\bf k}$.}
This amplitude had been  used for physical applications in
the framework of the Kadyshevsky's version of the quasipotential
approach~\cite{Kadysh,Skachkov}.

On the other hand, in  ref.~\cite{Weinberg}
an attractive $2(2j+1)$ component formalism for describing
particles of higher spins has been proposed.
As opposed to the Proca 4-vector potentials which transform according
to the $({1\over 2}, {1\over 2})$ representation of
the Lorentz group, the $2(2j+1)$ component functions are constructed
via the representation $(j,0)\oplus (0,j)$ in the Weinberg formalism.
This description of higher spin particles is on an equal
footing to the description of the Dirac spinor particle,
whose field function transforms according to the
$({1\over 2},0)\oplus (0, {1\over 2})$ representation.
The $2(2j+1)$- component analogues of the Dirac functions in
the momentum space are
\begin{equation}\label{pos}
{\cal U} ({\bf p})= \sqrt{{M\over 2}} \left (\matrix{
D^J \left (\alpha({\bf p})\right )\xi_\sigma\cr
D^J \left (\alpha^{-1\,\dagger}({\bf p})\right )\xi_\sigma\cr
}\right )\quad,
\end{equation}
for the positive-energy states; and\footnote{When setting ${\cal V} ({\bf p}) =
S^c_{[1]} \,{\cal U} ({\bf p}) \,\equiv \,{\cal C}_{[1]} \,
{\cal K}\, {\cal U} ({\bf p}) \,\sim \,\gamma_5 {\cal U} ({\bf p})$,
like the Dirac $j=1/2$ case we have other type of theories~\cite{Wigner,Sankar,Ah}.  $S^c_{[1]}$ is the
charge conjugation operator for $j=1$. ${\cal K}$ is the
operation of complex conjugation. }
\begin{equation}\label{neg}
{\cal V} ({\bf p})= \sqrt{{M\over 2}} \left (\matrix{
D^J \left (\alpha({\bf p})\Theta_{[1/2]}\right )\xi^*_\sigma\cr
D^J \left ( \alpha^{-1\,\dagger}({\bf p}) \Theta_{[1/2]}\right
)(-1)^{2J}\xi^*_\sigma\cr }\right )\quad,
\end{equation} for the
negative-energy states, ref.~\cite[p.107]{Novozh}, with the following
notations being used:
\begin{equation}
\alpha({\bf p})=\frac{p_0+M+(\bf{\sigma}
\cdot{\bf p})}{\sqrt{2M(p_0+M)}},\quad
\Theta_{[1/2]}=-i\sigma_2\quad.
\end{equation}
These functions obey
the orthonormalization equations,
${\cal U}^\dagger ({\bf p})\gamma_{00}\,{\cal U} ({\bf p})= M $,
$M$ is the mass of
the $2(2j+1)-$ particle. The similar normalization condition
exists for  $ {\cal V} ({\bf p})$, the functions of
``negative-energy states".

For instance, in the case of spin $j=1$, one has
\begin{eqnarray}
&&D^{\,1}\left (\alpha({\bf p})\right ) \,=\,
1+\frac{({\bf J}\cdot{\bf p})}{M}+
\frac{({\bf J}\cdot{\bf p})^2}{M(p_0+M)}\quad,\\
&&D^{\,1}\left (\alpha^{-1\,\dagger}({\bf p})\right ) \,=\,
1-\frac{({\bf J}\cdot{\bf p})}{M}+
\frac{({\bf J}\cdot{\bf p})^2}{M(p_0+M)}\quad,  \\
&&D^{\,1}\left (\alpha({\bf p}) \Theta_{[1/2]}\right ) \,=\,
\left [1+\frac{({\bf J}\cdot{\bf p})}{M}+
\frac{({\bf J}\cdot{\bf p})^2}{M(p_0+M)}\right ]\Theta_{[1]}\quad, \\
&&D^{\,1}\left (\alpha^{-1\,\dagger}({\bf p}) \Theta_{[1/2]}\right ) \,=\,
\left [1-\frac{({\bf J}\cdot{\bf p})}{M}+
\frac{({\bf J}\cdot{\bf p})^2}{M(p_0+M)}\right ]\Theta_{[1]}\quad,
\end{eqnarray}
($\Theta_{[1/2]}$,\,$\Theta_{[1]}$ are the Wigner operators for spin 1/2
and 1, respectively). Recently, much attention has been paid to this
formalism~\cite{DVO0}.   

In refs.~\cite{Novozh,Weinberg,Hammer,Dvoegl,Dvoegl1}
the Feynman diagram technique was
discussed in the above-mentioned six-component
formalism for particles of spin $j=1$. The Lagrangian is the following 
one:\footnote{In the following I prefer to use
the Euclidean metric because this metric got application in
a lot of papers on the $2(2j+1)$ formalism.}
\begin{eqnarray}
\lefteqn{{\cal L}=  \nabla_\mu \overline{\Psi}(x)\Gamma_{\mu\nu}
\nabla_\nu\Psi(x) - M^2\overline{\Psi}(x)\Psi(x)-{1 \over
4}F_{\mu\nu}F_{\mu\nu}+}\nonumber\\
&+&\frac{e\lambda}{12}F_{\mu\nu}\overline{\Psi}(x)\gamma_{5,\mu\nu}
\Psi(x)+\frac{e\kappa}{12 M^2}\partial_{\alpha}F_{\mu\nu}\overline{\Psi}(x)
\gamma_{6,\mu\nu,\alpha\beta}\nabla_{\beta}\Psi(x)\,.
\end{eqnarray}
In the above formula we have
$\nabla_{\mu}=-i\partial_{\mu}\mp eA_{\mu}$;
$F_{\mu\nu}=\partial_{\mu}A_{\nu}-\partial_{\nu}A_{\mu}$ is the
 electromagnetic field tensor; $A_{\mu}$ is the 4-vector of
electromagnetic field; $\overline{\Psi}, \Psi$  are
the six-component field functions
of the massive $j=1$ Weinberg particle.
The following expression has been obtained
for the interaction vertex of the particle with the vector potential, ref.~\cite{Hammer,Dvoegl}:
\begin{equation}
-e\Gamma_{\alpha\beta}(p+k)_{\beta} - {ie\lambda \over
6}\gamma_{5,\alpha\beta}q_{\beta}+{e\kappa \over
6M^2}\gamma_{6,\alpha\beta,\mu\nu}q_{\beta}q_{\mu}(p+k)_{\nu}\quad,
\label{22}
\end{equation}
where $\Gamma_{\alpha\beta}=\gamma_{\alpha\beta}+\delta_{\alpha\beta}$;\,
$\gamma_{\alpha\beta}$; \,$\gamma_{5, \alpha\beta}$; \,
$\gamma_{6,\alpha\beta, \mu\nu}$ \, are the $6\otimes 6$-matrices which
have been  described in ref.~\cite{Barut,Weinberg}:
\begin{eqnarray}
\gamma_{ij}\,&=&\,\pmatrix{
0 & \delta_{ij}\openone - J_i J_j- J_j J_i \cr
\delta_{ij}\openone - J_i J_j- J_j J_i & 0 \cr
}\quad,\\
\gamma_{i4}\,&=&\,\gamma_{4i}=\pmatrix{
0 & iJ_i \cr
-iJ_i & 0 \cr
}\quad,\quad
\gamma_{44}=\pmatrix{
0 & \openone \cr
\openone & 0\cr
}\quad,
\end{eqnarray}
and
\begin{eqnarray}
\gamma_{5,\alpha\beta}&=&i [\gamma_{\alpha\mu}, \gamma_{\beta\mu}]_-\quad,\\
\gamma_{6,\alpha\beta,\mu\nu}&=&
[\gamma_{\alpha\mu},\gamma_{\beta\nu}]_{+}
+2\delta_{\alpha\mu}\delta_{\beta\nu}-[\gamma_{\beta\mu},
\gamma_{\alpha\nu}]_{+}-2\delta_{\beta\mu}\delta_{\alpha\nu}\quad.
\end{eqnarray}
$J_i$ are the  spin matrices for a $j=1$  particle,
$e$ is  the electron charge, $\lambda$ and $\kappa$ 
correspond to the magnetic dipole moment and the electric quadrupole moment,
respectively.

In order to obtain the 4-vector current for the interaction
of a boson with the external field
one can use the known formulas of refs.~\cite{Skachkov,Chesh}, which
are valid for any spin:
\begin{equation}
{\cal U}^\sigma({\bf p}) =
\bf{S}_{{\bf p}} \,{\cal U}^\sigma({\bf 0})\quad, \quad \bf{S}_{{\bf
p}}^{-1} \bf{S}_{{\bf k}} = \bf{S}_{{\bf k}(-){\bf p}}\cdot I\otimes
D^{1}\left \{ V^{-1}(\Lambda_p, k)\right \}\quad,
\end{equation}
\begin{equation}
W_\mu({\bf p})\cdot D\left \{ V^{-1}(\Lambda_{p}, k)\right \}
= D\left \{ V^{-1}(\Lambda_{p}, k)\right \}
\cdot\left [ W_\mu({\bf k})
+\frac{p_\mu+k_\mu}{M(\Delta_0+M)}p_\nu W_\nu ({\bf k})\right ],
\end{equation}
\begin{equation}
k_\mu W_\mu ({\bf p})\cdot D\left \{ V^{-1}(\Lambda_{p}, k)\right \} =
-D\left \{ V^{-1}(\Lambda_{p}, k)\right \}\cdot p_\mu W_\mu ({\bf k})
\quad.
\end{equation}
$W_\mu$ is the Pauli-Lubanski 4-vector of relativistic spin.\footnote{It is usually 
introduced because 
the usual commutation relation for spin is not covariant in the relativistic domain.
The Pauli-Lubanski 4-vector is defined as
\begin{equation}
W_\mu ({\bf p}) = ( \Lambda_{\bf p} )_\mu^\nu W_\nu ({\bf 0})\,,
\end{equation}
where $W_0 ({\bf 0}) =0$, ${\bf W} ({\bf 0}) = M\bf{\sigma} /2$.
The properties are:
\begin{equation}
p^\mu W_\mu ({\bf p}) =0\,,\quad W^\mu ({\bf p}) W_\mu ({\bf p}) =-M^2 j (j+1)\,.
\end{equation}
The explicit form is
\begin{equation}
W_0 ({\bf p}) = ({\bf S}\cdot {\bf p})\,,\quad
{\bf W} ({\bf p}) = M{\bf S} + \frac{{\bf p} ({\bf S}\cdot {\bf p})}{p_0 +M}\,.
\end{equation}
}
The matrix $D^{(j=1)} \left \{ V^{-1} (\Lambda_{p}, k)\right \}$
is for spin 1:
\begin{eqnarray}
\lefteqn{D^{(j=1)}\left \{ V^{-1}(\Lambda_{p}, k)\right \}=
\frac{1}{2M(p_0+M)(k_0+M)(\Delta_0+M)} \left \{
\left [{\bf p}\times {\bf k}\right ]^2+\right.}\nonumber\\
&+&\left.\left [(p_0+M)(k_0+M) -{\bf k}\cdot{\bf p}\right ]^2 +
2i\left [(p_0+M)(k_0+M)-{\bf k}\cdot{\bf p}\right ] \left \{
{\bf J}\cdot\left [{\bf p}\times {\bf k}\right ]\right \}-\right.\nonumber\\
&-&\left.2\{{\bf J}\cdot\left
[{\bf p}\times{\bf k}\right ]\}^2\right \}\quad.
\end{eqnarray}
The formulas have been obtained in
ref.~\cite{Dvoegl1}:
\begin{eqnarray}
\bf{S}_{{\bf p}}^{-1} \gamma_{\mu\nu} \bf{S}_{{\bf p}}\,
&=& \,\gamma_{44} \left \{ \delta_{\mu\nu} - {1\over M^2}
\chi_{[\mu\nu]} ({\bf p})\otimes \gamma_5 - {2\over M^2}
\Sigma_{[\mu\nu]} ({\bf p})\right \}\quad,\\
\bf{S}_{{\bf p}}^{-1} \gamma_{5,\mu\nu} \bf{S}_{{\bf p}}\,
&=&\, 6i \left \{ - {1\over M^2}
\chi_{(\mu\nu)} ({\bf p})\otimes \gamma_5 + {2\over M^2}
\Sigma_{(\mu\nu)} ({\bf p})\right \}\quad,
\end{eqnarray}
where
\begin{eqnarray}
\chi_{[\mu\nu]} ({\bf p}) \,&=&\, p_\mu W_\nu ({\bf p})
+ p_\nu W_\mu ({\bf p})\quad,\\
\chi_{(\mu\nu)} ({\bf p}) \, &=&\, p_\mu W_\nu ({\bf p})
- p_\nu W_\mu ({\bf p})\quad, \\
\Sigma_{[\mu\nu]} ({\bf p}) \,&=&\,{1\over 2}
\left \{ W_\mu ({\bf p}) W_\nu ({\bf p}) +
W_\nu ({\bf p}) W_\mu ({\bf p})\right \}\quad,\\
\Sigma_{(\mu\nu)} ({\bf p}) \,&=&\,{1\over 2}
\left \{ W_\mu ({\bf p}) W_\nu ({\bf p}) -
W_\nu ({\bf p}) W_\mu ({\bf p}) \right \}\quad,
\end{eqnarray}
lead to the 4- current of a $j=1$ Weinberg
particle more directly:\footnote{{\it Cf.} with a $j=1/2$
case:
\begin{eqnarray}
&&\bf{S}_{\bf p}^{-1} \gamma_\mu \bf{S}_{\bf k} = \bf{S}_p^{-1} \gamma_\mu \bf{S}_p \bf{S}_{{\bf k} (-) {\bf p}} I \otimes D^{1/2} \{ V^{-1} (\Lambda_{\bf p}, {\bf k})\} \,, \\
&&\bf{S}_p^{-1} \gamma_\mu \bf{S}_p = {1 \over m}
\gamma_0 \left \{ \openone \otimes p_\mu + 2\gamma_5 \otimes W_\mu ({\bf
p}) \right \} \,, \\
&&\bf{S}_p^{-1} \sigma_{\mu\nu} \bf{S}_p =
- {4\over m^2} \openone \otimes \Sigma_{(\mu\nu)} ({\bf p})
+ {2\over m^2} \gamma_5 \otimes \chi_{(\mu\nu)}
({\bf p})\,.
\end{eqnarray}
Of course, the product of two Lorentz boosts is {\it not}  a pure Lorentz transformation. It contains the rotation, which describes the Thomas spin precession (the Wigner rotation $V(\Lambda_{\bf p}, {\bf k}) \in SU(2))$). And, then,
\begin{equation}
j_\mu^{\sigma_p\nu_p} ({\bf k} (-) {\bf p}, {\bf p}) =
{1\over m} \xi^\dagger_{\sigma_p} \left \{2g_v \mbox{\ae}_0 p_\mu
+ f_v \mbox{\ae}_0 q_\mu + 4 g_{{\cal M}}  W_\mu ({\bf p})
(\bf{\sigma} \cdot \mbox{{\bf \ae}})\right \} \xi_{\nu_p}\quad,
\quad (g_{{\cal M}} = g_v + f_v)\,.\label{current}
\end{equation}
The indices ${\bf p}$ indicate that  the Wigner rotations have been separated
out and, thus,  all spin indices have been ``resetted"
on the momentum ${\bf p}$. One can re-write~[2b] the
electromagnetic current~(\ref{current}):
\begin{equation}
j_\mu^{\sigma_p\nu_p} ({\bf k}, {\bf p}) =
- {e\,m\over \mbox{\ae}_0} \xi^\dagger_{\sigma_p} \left \{
g_{{\cal E}} (q^2)\, (p+k)^\mu +
g_{{\cal M}} (q^2)\,
\left [{1\over m} W_\mu ({\bf p}) (\bf{\sigma}\cdot \bf{\Delta})
- {1\over m} (\bf{\sigma}\cdot \bf{\Delta}) W_\mu ({\bf p})\right ]
\right \} \xi_{\nu_p}\,.\label{current1}
\end{equation}
$g_{{\cal E}}$ and $g_{{\cal M}}$ are the analogues of the Sachs electric
and magnetic form factors.
Thus, if we regard $g_{S,T,V}$ as  effective coupling
constants depending on the momentum transfer one can ensure ourselves
that the forms of the currents for a spinor particle  and those
for a $j=1$ boson are the same (with the Wigner rotations separated out).}
\begin{eqnarray}
j_{\mu}^{\sigma_{p}\nu_{p}}({\bf p}, {\bf k}) &=&
j_{\mu \,(S)}^{\sigma_{p}\nu_{p}}({\bf p}, {\bf k}) +
j_{\mu \,(V)}^{\sigma_{p}\nu_{p}}({\bf p}, {\bf k}) +
j_{\mu \,(T)}^{\sigma_{p}\nu_{p}}({\bf p}, {\bf k})\quad,\\
j_{\mu \,(S)}^{\sigma_{p}\nu_{p}}({\bf p}, {\bf k}) \,&=&\,
-\,g_S \xi^\dagger_{\sigma_p} \left \{  (p+k)_\mu \left (
1+ \frac{({\bf J}\cdot \bf{\Delta})^2}{M (\Delta_0 + M)} \right )\right
\} \xi_{\nu_p}\quad,\label{curs}\\
j_{\mu \,(V)}^{\sigma_{p}\nu_{p}}({\bf
p}, {\bf k}) \,&=&\, -\,g_V \xi^\dagger_{\sigma_p} \left \{ (p+k)_{\mu}+
{1\over M}W_{\mu}({\bf p})({\bf J}\cdot\bf{\Delta})- {1\over M}({\bf
J}\cdot\bf{\Delta}) W_{\mu}({\bf p})\right \} \xi_{\nu_p}\quad,
\label{cur}\\
j_{\mu \,(T)}^{\sigma_{p}\nu_{p}}({\bf p}, {\bf k}) \,&=&\,
-\, g_T \xi_{\sigma_p}^\dagger \left \{ - (p+k)_\mu
\frac{({\bf J}\cdot \bf{\Delta})^2}{M (\Delta_0 + M)}+
\right.\label{curt}\\
&& \left. \qquad\qquad\qquad + {1\over M} W_{\mu}({\bf p})({\bf
J}\cdot\bf{\Delta})- {1\over M}({\bf J}\cdot\bf{\Delta}) W_{\mu}({\bf
p})\right \} \xi_{\nu_p}\quad.\nonumber
\end{eqnarray}
Next, let me  now present the Feynman matrix
element corresponding to the diagram of  two-boson interaction, mediated
by the particle described by the vector
potential,  in the form~\cite{Skachkov,Dvoegl} (read the remark in
the footnote \# 1):
\begin{eqnarray}
\lefteqn{ <p_1, p_2;\, \sigma_1, \sigma_2\vert \hat T^{(2)}
\vert k_1, k_2;\, \nu_1, \nu_2>
=}\nonumber\\
&=&\sum^{1}_{\sigma_{ip}, \nu_{ip}, \nu_{ik} =-1} D^{\dagger\quad
(j=1)}_{\sigma_1\sigma_{1p}} \left \{V^{- 1} (\Lambda_{\cal P}, p_1)\right
\} D^{\dagger\quad (j=1)}_{\sigma_2\sigma_{2p}} \left \{V^{-1}
(\Lambda_{\cal P}, p_2)\right \}\times\nonumber\\
&\times&T^{\nu_{1p}\nu_{2p}}_{\sigma_{1p}\sigma_{ 2p}}({\bf k} (-)
{\bf p}, {\bf  p}) D^{(j=1)}_{\nu_{1p}\nu_{1k}}\left \{V^{-1}
(\Lambda_{p_1}, k_1)\right \} D^{(j=1)}_{\nu_{1k}\nu_1}\left \{V^{-1}
(\Lambda_{\cal P}, k_1)\right \}\times\nonumber\\
&\times& D^{(j=1)}_{\nu_{2p}\nu_{2k}} \left\{ V^{-1} (\Lambda_{p_2},
k_2)\right \} D^{(j=1)}_{\nu_{2k}\nu_2}\left\{ V^{-1} (\Lambda_{\cal P},
k_2)\right \}\quad,
\end{eqnarray}
where
\begin{equation} \label{ampl}
T^{\nu_{1p}\nu_{2p}}_{\sigma_{1p}\sigma_{2p}} ({\bf k}(-) {\bf p}, {\bf p}) =
\xi^\dagger_{\sigma_{1p}} \xi^\dagger_{\sigma_{2p}} \,
T^{(2)} ({\bf k} (-) {\bf p},\, {\bf p})\, \xi_{\nu_{1p}} \xi_{\nu_{2p}}\quad,
\end{equation}
$\xi^\dagger$, $\xi$ are the 3-analogues of 2-spinors.
The calculation of the amplitude (\ref{ampl}) yields
($p_0 = -ip_4$,\,\,$\Delta_0 = -i \Delta_4$):
\begin{eqnarray}\label{212}
\lefteqn{\hat T^{(2)} ({\bf k}(-){\bf p}, {\bf p})
\,=\, g^2 \left\{ \frac{\left [p_0
(\Delta_0 +M) + ({\bf p}\cdot \bf{\Delta})\right ]^2
-M^3 (\Delta_0+M)}{M^3 (\Delta_0 -M)}+\right.}\nonumber\\
&+&\left.\frac{i ({\bf J}_1+{\bf J}_2)\cdot\left [{\bf p}
\times\bf{\Delta}\right ]}
{\Delta_0-M}\left [ \frac{p_0 (\Delta_0 +M)+{\bf p}\cdot
\bf{\Delta}}{M^3} \right ]
+ \frac{({\bf J}_1\cdot \bf{\Delta})({\bf
J}_2\cdot \bf{\Delta})-({\bf J}_1\cdot{\bf J}_2) \bf{\Delta}^2}{2M
(\Delta_0-M)}-\right.\nonumber\\
&-&\left.\frac{1}{M^3}\frac{{\bf J}_1\cdot\left [{\bf p}
\times\bf{\Delta}\right ] \,\,{\bf J}_2\cdot \left [{\bf
p} \times\bf{\Delta}\right ]}{\Delta_0-M}\right\}\quad.
\end{eqnarray}
We have assumed $g_S = g_V = g_T$ above. The expression (\ref{212}) reveals  the
advantages of the $2(2j+1)$- formalism, since
it looks like  the amplitude for the interaction of two spinor particles
with the substitutions
$$\frac{1}{2M(\Delta_0 - M)}
\Rightarrow\frac{1}{\bf{\Delta}^2}\quad \mbox{and}
\quad {\bf J}\Rightarrow \bf{\sigma}\quad.$$
The calculations hint that many analytical results produced for
a Dirac fermion could be applicable to describing a $2(2j+1)$
particle. Nevertheless, an adequate explanation is required
for the obtained difference.   You may see that
\begin{equation}
{1\over {\bf \Delta}^2} = {1\over 2M (\Delta_0 - M)} - {1\over 2M (\Delta_0 +M)}
\end{equation}
and 
\begin{equation}
(p+k)_\mu (p+k)^\mu = 2M (\Delta_0 +M)\,.
\end{equation}
Hence, if we add an additional diagramm of another channel (${\bf k} \rightarrow -{\bf k}$), we can obtain the {\it full}
coincidence in the $T$-matrices of the fermion-fermion interaction and the boson-boson interaction. But, of course, one should take into account that  there is no the Pauli principle for bosons, and additional sign $``-"$ would be related to the indefinite metric.

So, the conclusions are:
The main result of this paper is the boson-boson amplitude
calculated in the framework of the $2(2j+1)-$ component theory.
The separation of the Wigner rotations permits us to reveal
certain similarities with the  $j=1/2$ case. Thus, this result
provides a ground for the conclusion: if
we would accept the description of higher spin particles 
on using the Weinberg $2(2j+1)-$ scheme many calculations produced 
earlier for fermion-fermion
interactions mediated by the vector potential can be applicable
to processes involving higher-spin particles. Moreover,
the main result of the paper gives a certain hope at a possibility
of the unified description of fermions and bosons.
One should realize that all the above-mentioned is
not surprising.  The principal features of describing a particle 
on the basis of relativistic quantum field theory are {\it not} in some
special representation of the group representation, $(1/2,0)\oplus (0,1/2)$,
or $(1,0)\oplus (0,1)$, or $(1/2,1/2)$, but in the Lorentz group itself.
However, certain differences between denominators
of the amplitudes are still not explained in full.

Several works dealing with  phenomenological description
of hadrons in the $(j,0)\oplus (0,j)$ framework have
been published~\cite{DVO-old2,DVO-old3,DVO-pr}.

\ack 
This paper is based on the talks given at the 5th International Symposium on ``{\it Quantum Theory and Symmetries}", July 22-28, 2007, Valladolid, Spain and the 10th Workshop ``{\it What comes beyond the Standard Model?}", July 17-27, 2007, Bled, Slovenia.
I am grateful to participants of recent conferences for discussions.

\end{document}